# Broadband asymmetric transmission with tunable bilayer silicon nanoarrays: from visible to near-infrared

Ruihan Ma[1], Yuqing Cheng[1,*] and Mengtao Sun[1,†]

[1] School of Mathematics and Physics, University of Science and Technology Beijing, Beijing 100083, People's Republic of China

**Abstract:** A kind of asymmetric transmission (AT) device based on bilayer silicon arrays (BSA) nanostructure is theoretically explored, which achieves high forward transmissivity and suppressed backward transmissivity for broadband by simply adjusting the parameters of the structure. The structure consists of two silicon cylinder arrays, one on the $SiO_2$ substrate and the other embedded in the substrate. Particularly, three AT devices with different configurations are designed, which exhibit broadband AT with high isolation ratios in the wavelength ranges of 685-807 nm, 866-1029 nm, and 1285-1536 nm, respectively. A comprehensive analysis of the BSA structure's performance across different array periods highlights its potential for broadband optical applications, such as optical isolation and multi-channel optical sensors.

## 1. Introduction

With the rapid development of photonic integration technology, achieving efficient and controllable light propagation within on-chip platforms has become an important research area in nanophotonics[1-4]. AT devices offer distinct optical transmissivities for forward and backward propagation[5-12], offering an effective approach for directional control of optical signals. Such devices have attracted considerable interest due to their potential applications in optical isolation[13-16], information encryption[17, 18], neuromorphic optical computing[19, 20], and optical communication systems[21-23]. For broadband information processing and multi-band optical systems, AT devices that maintain high forward transmissivity and effective backward suppression over a wide wavelength range are essential. Wei et al. designed an all-dielectric AT structure composed of a nanograting coupled with a defective multilayer photonic crystal, achieving multiband and bidirectional multiplexed AT in the visible region, with a maximum forward-backward transmissivity difference of 0.974, demonstrating excellent directional selectivity[24]. Yang et al. proposed an AT device based on periodic silicon (Si) cylinder arrays on a dielectric substrate, realizing efficient and polarization-independent AT over a broad wavelength range from 300 to 1100 nm, where the forward transmissivity approaches 0.9 at 540 nm while the backward transmission is almost completely suppressed[25]. Cheng et al. reported a hybrid metallic nanowaveguide in which asymmetric arrays were introduced at both the input and output sides, enabling multimode AT in the visible region[26].

[*] Email: yuqingcheng@ustb.edu.cn
[†] Email: mengtaosun@ustb.edu.cn

In this study, we employ a bilayer Si array (BSA) nanostructure, where the top layer is a Si cylinder array on the $SiO_2$ substrate, and the bottom layer is a Si cylinder array embedded in the substrate. By tuning the geometric parameters of the BSA, high transmissivity for broadband can be achieved at different wavelength regions. This compact and feasible structure provides new physical insight into multi-mode coupling mechanisms in asymmetric nanostructures.

## 2. Method and structure

The schematic in Fig. 1 shows periodic Si cylinder array on a $SiO_2$ substrate. The structure comprises an upper array and a lower array. Each unit cell of the upper array contains a single silicon cylinder, whereas the lower array consists of four silicon cylinders embedded in the $SiO_2$ substrate and arranged periodically. This results in the period of the upper two being times the period of the lower. Tuning the cylinder radius (r), height (h), and array period (p) yields high forward transmissivity and suppressed backward transmissivity across different wavelength bands, enabling efficient AT.

Three sets of optimized structural parameters corresponding to three distinct operating wavelengths are investigated in this work, as summarized in Table 1. The interface between the silicon cylinders and the substrate is defined as z = 0.

The optical response of the device is numerically investigated using the three-dimensional finite-difference time-domain (FDTD) method[27-29]. The incident light is an x-polarized plane wave propagating along the −z (forward) and the z (backward) direction, with the incident angle set to 0° unless otherwise specified. Periodic boundary conditions are applied along the x and y directions, while perfectly matched layers (PMLs) are employed along the ±z boundaries to absorb outgoing radiation and emulate an open-space environment. The refractive indices of the constituent materials are taken from Refs.[30, 31]. To quantitatively evaluate the optical AT performance, the isolation ratio (IR) of transmissivity, expressed in the unit of dB, is defined as

$$\text{IR} = 10 \times \log_{10}\left(\frac{T_f}{T_b}\right) \quad (1)$$

where $T_f$ and $T_b$ denote the forward and backward transmittances, respectively. The IR is used to quantify the transmission contrast between forward and backward illumination.

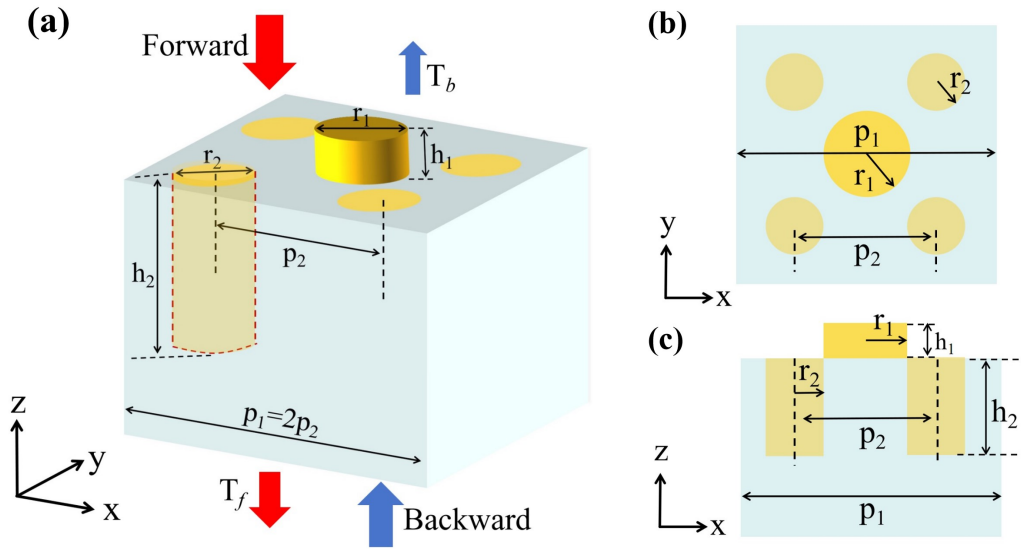

**Fig. 1.** Schematic of the unit cell for the periodic nanostructure. (a) 3D view. (b) Top view (xy

plane). (c) Front view (xz plane). The yellow and light-gray regions represent Si and SiO$_2$, respectively.

Table 1. Parameters of the AT Devices

| Parameters | unit: nm | | |
|---|---|---|---|
| Upper array period ($p_1$) | 600 | 1600 | 2400 |
| Lower array period ($p_2$) | 300 | 800 | 1200 |
| Upper cylinder height ($h_1$) | 120 | 150 | 50 |
| Lower cylinder height ($h_2$) | 220 | 290 | 430 |
| Upper cylinder radius ($r_1$) | 110 | 100 | 200 |
| Lower cylinder radius ($r_2$) | 110 | 300 | 470 |
| working band | 685-807 | 866-1029 | 1285-1536 |

## 3. Results and discussion

*3.1 Asymmetric spectra*

Fig. 2 shows the forward and backward transmission spectra of the three device with different parameters, marked with upper array periods $p_1$ = 600, 1600, and 2400 nm (Figs. 2**(a)-(c)**) along with the corresponding IR (Figs. 2**(d)-(f)**). These results are used to analyze how the structure achieves high transmission contrast at different wavelengths across a wide bandwidth by varying the parameters.

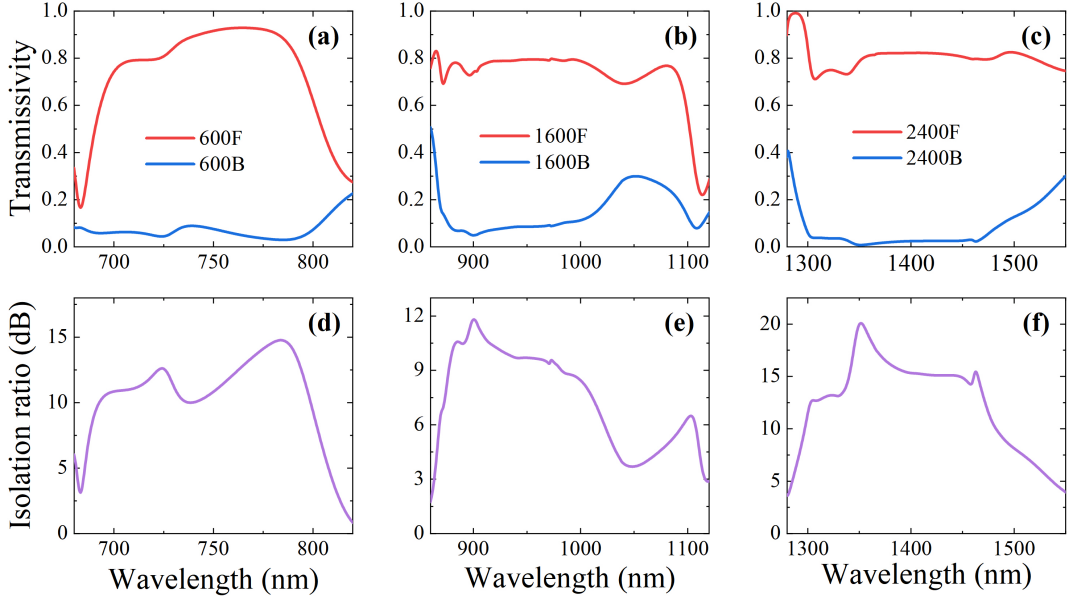

**Fig. 2.** (a)-(c) Transmission spectra of the three devices for $p_1$ = 600, 1600, and 2400 nm, respectively. Red and blue curves stand for forward (F) and backward (B) transmissivities, respectively. (d)-(f) IR for $p_1$ = 600, 1600, and 2400 nm, respectively.

As shown in Fig. 2**(a)**, when the array period $p_1$ = 600 nm, the forward transmissivity is larger than 0.7 for a broadband (695-796 nm) with a maximum of 0.93, while the backward transmissivity is less than 0.15 for broadband (680-810 nm) with a minimum of around 0.05. This results in a significant broadband AT. As the array period increases, the wavelength ranges with high IR redshift. Particularly, at $p_1$ = 1600 nm, the forward transmissivity is larger than 0.7 for a broadband (873-1033 nm) with a maximum of 0.79, while the backward transmissivity is less than 0.15 for broadband (869-1015 nm) with a minimum of around 0.048. When the array period $p_1$ = 2400 nm, the forward transmissivity is larger than 0.7 for a broadband (1289-1905 nm) with a maximum of 0.82, while the backward transmissivity is less than 0.15 for broadband (1292-1510 nm) with a minimum of around 0.007.

Figs. 2**(d)-(f)** show the IR for forward and backward transmission at different array periods. Here, we define the AT band and its bandwidth of the device as the wavelength range in which IR is larger than 5 dB. Therefore, the bandwidth of $p_1$ = 600 nm device is 122 nm, ranging from 685 to 807 nm, with the IR reaching a maximum of 14.8 dB. When the array period increases to $p_1$ = 1600 nm, the bandwidth is 163 nm, ranging from 866 to 1029 nm, with the IR reaching a maximum of 11.8 dB. When the array period increases to $p_1$ = 2400 nm, the bandwidth is 251 nm, ranging from 1285 to 1536 nm, with the IR reaching a maximum of 20.1 dB.

In addition, varying the polarization angle of the incident light does not lead to noticeable changes in the transmission characteristics of the devices (not shown), indicating that the structure exhibits excellent polarization insensitivity. This behavior can be attributed to the high in-plane symmetry of the structure, which results in identical optical responses to different linear polarization directions.

This section demonstrates the excellent broadband performance of the structure through the analysis of transmission characteristics at different array periods. The results show that as the array period increases, the IR remains high, confirming the advantage of the structure's broadband AT characteristics. Particularly across multiple wavelength ranges, the results indicate excellent optical AT performance over broad wavelength ranges, and highlighting its potential for practical applications optical filters and multi-channel optical sensors.

## 3.2 Electric field modes

We define $E_0$ = 1 V/m as the electric field intensity of the source. Figs. 3 and 4 show the electric field distributions ($E_x$, $E_y$, and $E_z$ components) of the device for $p_1$ = 600 nm at the y = 150 nm and z = −110 nm planes at the wavelength of 774 nm, respectively.

Fig. 3 shows the electric field distributions of the device in the xz plane, including both forward and backward incidences. For forward incidence, Fig. 3**(a)** shows the $E_x$ component, where the electric field varies strongly along the x-axis, with significant oscillations near the central region of the structure, indicating that the incident light is effectively coupled into the structure. Fig. 3**(b)** shows the $E_y$ component, where the electric field is mainly distributed near the lower cylinders. Fig. 3**(c)** shows the $E_z$ component, where the electric field exhibits strong localization near interface and the surfaces of the lower cylinders, indicating the effective propagation and transmission of light within the structure. For backward incidence, Fig. 3(d)-(f) show the electric field distribution. Compared to forward incidence, the field strength is significantly lower in backward incidence, indicating that backward transmission is suppressed.

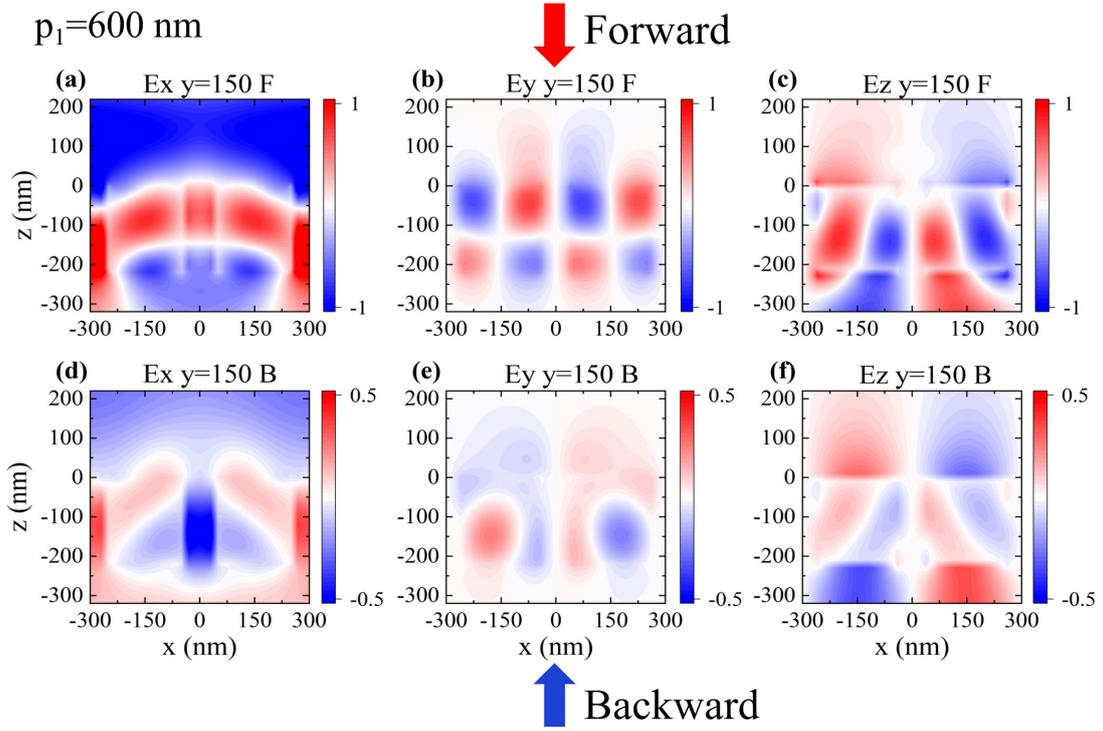

**Fig. 3.** The electric field distributions of the device with $p_1$ = 600 nm at the y = 150 nm plane at the wavelength of 774 nm. (a-c) Forward incidence: (a) $E_x$. (b) $E_y$. (c) $E_z$. (d-f) Backward incidence: (d) $E_x$. (e) $E_y$. (f) $E_z$. The color bar represents the electric field intensity.

Fig. 4 shows the electric field distributions of the device in the xy plane, including both forward and backward incidences. For forward incidence, Fig. 4(a) shows the $E_x$ component, where the electric field is strongly enhanced in the central region of the structure, indicating strong coupling of the incident light in this direction. Fig. 4(b) shows the $E_y$ component, where the electric field is cylindrical symmetric about the z axis. Fig. 4(c) shows the $E_z$ component, where the electric field exhibits significant localization near the surface of the lower cylinders. For backward incidence, Fig. 4(d)-(f) show the electric field distribution. Compared to forward incidence, the field strength is weaker in backward incidence, further verifying the suppression of backward transmission.

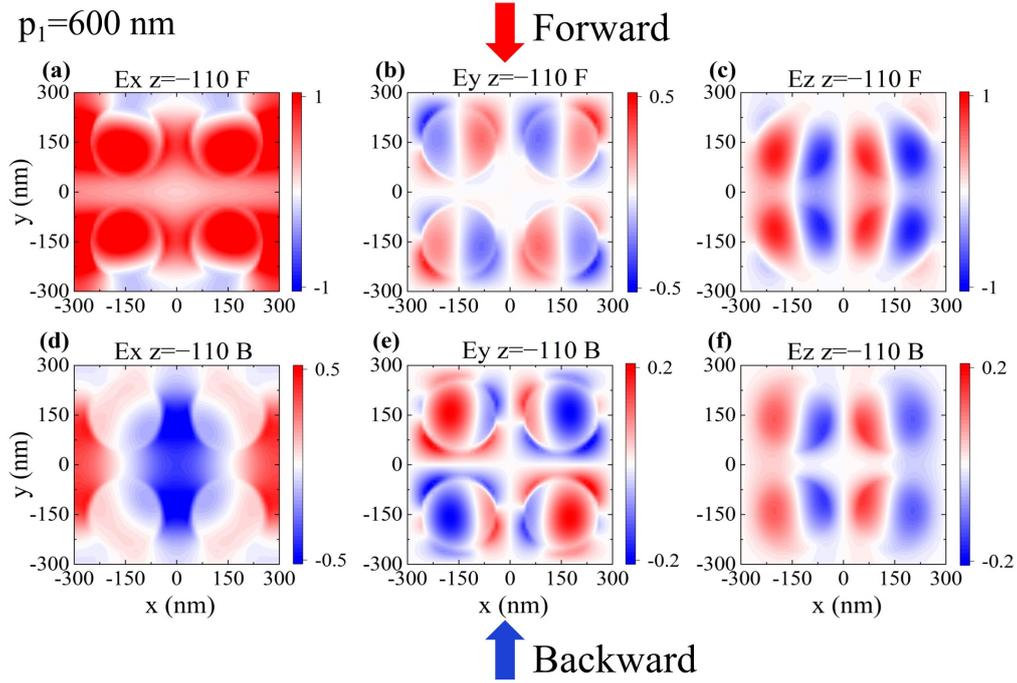

**Fig. 4.** The electric field distributions of the device with $p_1 = 600$ nm at the $z = -110$ nm plane at the wavelength of 774 nm. (a-c) Forward incidence: (a) $E_x$. (b) $E_y$. (c) $E_z$. (d-f) Backward incidence: (d) $E_x$. (e) $E_y$. (f) $E_z$. The color bar represents the electric field intensity.

Fig. 3 and Fig. 4 show the electric field distributions of the device in the xz plane and xy plane, respectively. Although these two planes provide different perspectives, they complement each other and together reveal the propagation characteristics of light within the structure. In the xz plane (Fig. 3), the distribution of the electric field along the x-axis shows how the forward incident light is effectively coupled into the structure, while in the xy plane (Fig. 4), the distribution of the electric field further validates the effect of the array coupling on the propagation of the electric field. By comprehensively analyzing the electric field distributions in these two planes, we can gain a more comprehensive understanding of the field localization effects in the structure and their contribution to the forward transmission.

To fully demonstrate the impact of different array periods on the electric field distribution, the electric field distributions in the xz and xy planes for $p_1 = 1600$ nm and $p_1 = 2400$ nm are shown in Figs. S1-S4 in Supplementary Materials. The results indicate that the electric field distributions for these array periods are similar to those for $p_1$ = 600 nm, further confirming the AT behavior of the structure at different periods.

## 4. Conclusion

In summary, the AT device based on the BSA structure demonstrates excellent AT performance, with high forward and low backward transmissivities across wide ranges of wavelengths. By tuning the parameters of the structure, three typical AT devices are designed for different wavelength ranges, all of which exhibit broadband AT. For the device with the array period of $p_1$ = 600 nm, broadband AT is achieved over the wavelength range from 685 to 807 nm, corresponding to a bandwidth of 122 nm. The maximum IR reaches 14.8 dB at the wavelength of 784 nm. When the array period is increased to $p_1$ = 1600 nm, the AT band exhibits a clear redshift toward longer wavelengths. In this case, the wavelength ranges from 866 to 1029 nm, corresponding to a bandwidth of 163 nm. The maximum IR is 11.8 dB at the wavelength of 901 nm. When the array period is further increased to $p_1$ = 2400 nm, the device shows a more pronounced redshift effect for the AT band together with significant bandwidth enhancement, i.e., over the wavelength range from 1285 to 1536 nm, corresponding to a bandwidth of 251 nm, with the maximum IR reaching approximately 20.1 dB at the wavelength of 1350 nm.

These results confirm that the BSA structure can achieve high IR for broadbands. The multiple wavelength ranges of the three devices provide potential possibilities to extend the wavelength ranges to other bands. The results make it highly suitable for applications such as optical filters, multi-channel optical sensors, and optical isolation.

## Acknowledgments

This work was supported by the National Natural Science Foundation of China (Grant No. 12504461).

## Disclosures

The authors declare no conflicts of interest.

**Data availability statement**

The data that support the findings of this study are available upon reasonable request from the authors.